\documentclass[twocolumn]{aa}
\usepackage{txfonts}
\usepackage{graphicx}
\usepackage{lscape}
\begin{document}
\newcommand{\obj}{UM~673}
\title{Photometric monitoring of the doubly imaged quasar \obj:\\ 
\vspace*{1mm}
possible evidence  for chromatic microlensing
\thanks{Based on observations made with the Danish 1.5\,m telescope
 (ESO, La~Silla, Chile)
(Proposals:
63.O-0205(A), 64.O-0235(B),  
65.O-0214(B), 66.A-0203(B),
67.A-0115(B), 68.A-0109(B)).}}

\author{Th. Nakos \inst{1,2,3}
\and F. Courbin   \inst{4}
\and J. Poels     \inst{2}
\and C. Libbrecht \inst{2,3}
\and P. Magain    \inst{2}
\and J. Surdej    \inst{2}
\thanks{Directeur de recherche honoraire du
Fonds National de la Recherche Scientifique (Belgium)}
\and J. Manfroid  \inst{2}
\and I. Burud     \inst{5}
\and J. Hjorth    \inst{6}
\and L. Germany   \inst{1}
\and C. Lidman    \inst{1}
\and G. Meylan    \inst{4}
\and E. Pompei    \inst{1}
\and J. Pritchard \inst{7,6,1}
\and I. Saviane   \inst{1}}

\offprints{Th. Nakos, \email{nakos@astro.ulg.ac.be}}

\institute{
European Southern Observatory, Casilla 19001, Santiago, Chile
\and
Institut d'Astrophysique et de G\'eophysique, Universit\'e de Li\`ege, 
All\'ee du 6 Ao\^ut 17, Sart Tilman, B\^at. B5C, B-4000 Li\`ege, Belgium
\and 
Royal Observatory of Belgium, Ringlaan 3, B-1180 Brussels, Belgium      
\and
Laboratoire d'Astrophysique, Ecole Polytechnique F\'ed\'erale de
Lausanne (EPFL), Observatoire, CH-1290 Sauverny, Switzerland
\and
Space Telescope Science Institute, 3700 San Martin Drive, 
Baltimore, MD 21218, USA
\and 
Niels Bohr Institute, University of Copenhagen, Juliane Maries Vej 30, DK-2100, 
K{\o}benhavn {\O}, Denmark
\and
Mount John University Observatory and Department of Physics \& 
Astronomy, University of Canterbury, Private Bag 4800, Christchurch, New 
Zealand}

\date{Received / Accepted }

\abstract{We     present the results    of   two-band CCD  photometric
  monitoring   of  the   gravitationally  lensed  quasar Q\,0142$-$100
  (\obj).  The data, obtained  at ESO-La~Silla with the 1.54\,m Danish
  telescope  in the $Gunn\,i$-band (October  1998 $-$ September 1999)
  and  in the Johnson $V$-band  (October 1998 to  December 2001), were
  analyzed   using   three    different  photometric   methods.  The
  light-curves   obtained with  all  methods   show  variations, with  a
  peak-to-peak amplitude of  0.14 magnitude in  $V$.  Although it  was
  not possible to  measure the time delay between  the two lensed QSO images,
  the brighter component displays  possible evidence for microlensing:
  it becomes  bluer  as it  gets   brighter,  as expected   under  the
  assumption of differential magnification of a quasar accretion disk.

\keywords{Gravitational lensing  -- quasars: individual: Q\,0142$-$100
(\obj)}}

\authorrunning{Nakos et al.}
\titlerunning{Photometric monitoring of the lensed quasar \obj}
\maketitle

%________________________________________________________________

\section{Introduction}

\obj\   (Surdej  et    al.~\cite{surdej87})  is   one  of   the  first
gravitational lens systems (GLs) found in a  systematic  optical search. It
consists of two lensed quasar  images (z$_{\rm s}=2.719$) separated by
$2.2\arcsec$.  Although no   direct spectrum of  the  lensing galaxy has
ever been obtained,  CaII and NaI   absorption lines, detected  in the
spectrum of the  fainter quasar  image B,  suggest   that it  is at  a
redshift z$_{\rm  l}=0.49$ (Surdej  et al.~\cite{surdej88};  Smette et
al.~\cite{smette92}).   Because  of  its  relatively  bright  apparent
magnitude  (m$_{\rm   R}$(A)$\sim$16.5,  m$_{\rm R}$(B)$\sim$19.0) and
because of the wide angular separation between its  images, \obj\ is a
good target for   a  photometric monitoring program.  The   expected
time-delay is of the  order of a few  months, which is relatively easy
to measure, providing that   the  quasar itself displays   significant
photometric variations.

\obj\ has been  monitored at the  European  Southern Observatory (ESO,
La~Silla, Chile) from 1987 to 1993 with several telescopes, leading to
photometric measurements for 18 epochs. Unfortunately, the small field
of view (FOV) accessible with CCDs in the  eighties and the subsequent lack
of PSF stars, restricted   the effectiveness  of the observations   to
unveil any weak variation, at the level of a few tenths of a magnitude
(Daulie et al.~\cite{daulie93}).

%----------------------------------------------------

\begin{figure}[t]                                            
\centering
\includegraphics[width=8.7cm]{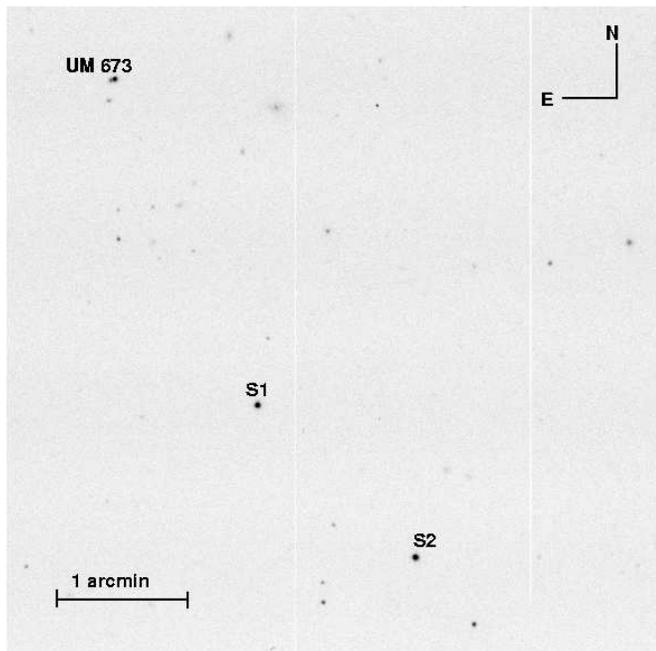}    
\caption{ Part of  one $Gunn\,i$ DFOSC CCD frame of  \obj.  The field
  of  view is $5\arcmin  \times  5\arcmin$.  Two  reference stars  are
  indicated (S1  and S2).  Star  S1, whose absolute photometry is well
  known (Nakos et al.  2003) has been used to normalize the CCD frames.  
  Star S2 was taken as a comparison star.
\label{findingchart}} 
\end{figure}

%----------------------------------------------------

Within the  framework   of the   Hamburg   Quasar  Monitoring  Program
(Borgeest  \& Schramm~\cite{borgest93}), \obj\  has also been observed
from 1988  to  1993 with  the 1.23\,m MPIA   telescope,  at Calar Alto
Observatory, Spain.  In contrast with Daulie et al.~(\cite{daulie93}),
a  \emph{preliminary} analysis led to the   detection of variations in
the light-curve of component A,  with a peak-to-peak amplitude of  the
order of 0.2 magnitude.

Sinachopoulos et  al.~(\cite{sina01}) reported variation  of up to 0.5
magnitude  over   several  years.    Nevertheless,  their  data   were
irregularly  distributed and the small   size of the telescopes  used,
under moderate seeing conditions,  only allowed the measurement of the
\emph{total} flux of the gravitational lens system.

The latest work on \obj\ (Wisotzki et al. ~\cite{wiso04}) reported on
spectra obtained in September  2002.   The data  were taken  with  the
3.5\,m  telescope  at  Calar  Alto     (Spain),  using   the   Potsdam
Multi-Aperture Spectrophotometer (PMAS).   By  measuring the  emission
line  to  continuum   flux  ratios for  the  two   quasar images, they
concluded   that, during the period  of  observations, no microlensing
signature was detectable.

\obj\  was part  of  a photometric  monitoring program carried  out at
ESO-La~Silla, using the  1.54\,m Danish telescope. In this monitoring,
the photometric variability of several GL systems was first tested for
a few months and the  object was either followed  up further in order
to  measure  the   time-delay    (see   for  example  Burud  et    al. 
~\cite{burud02a}), or   stopped   due  to  insufficient    photometric
variability.  \obj\  is  one of the   objects that was  stopped. 
However, it turned out to be interesting, after  a more detailed analysis
of the data was carried out.  Both components show variability on long
time-scales, with peak-to-peak  variations of  the  order of  0.14 and
0.08  magnitude,  in $V$  and  $i$ respectively.  While    further
(spectroscopic) observations  will be  needed to  discriminate between
intrinsic and microlensing induced variability, it  is argued, on the
basis of color information, that at  least part of the variability, on
time-scales of a few months, is likely due to microlensing.\\

%----------------------------------------------------

\begin{figure}[t]                                            
\centering
\includegraphics[width=7.cm]{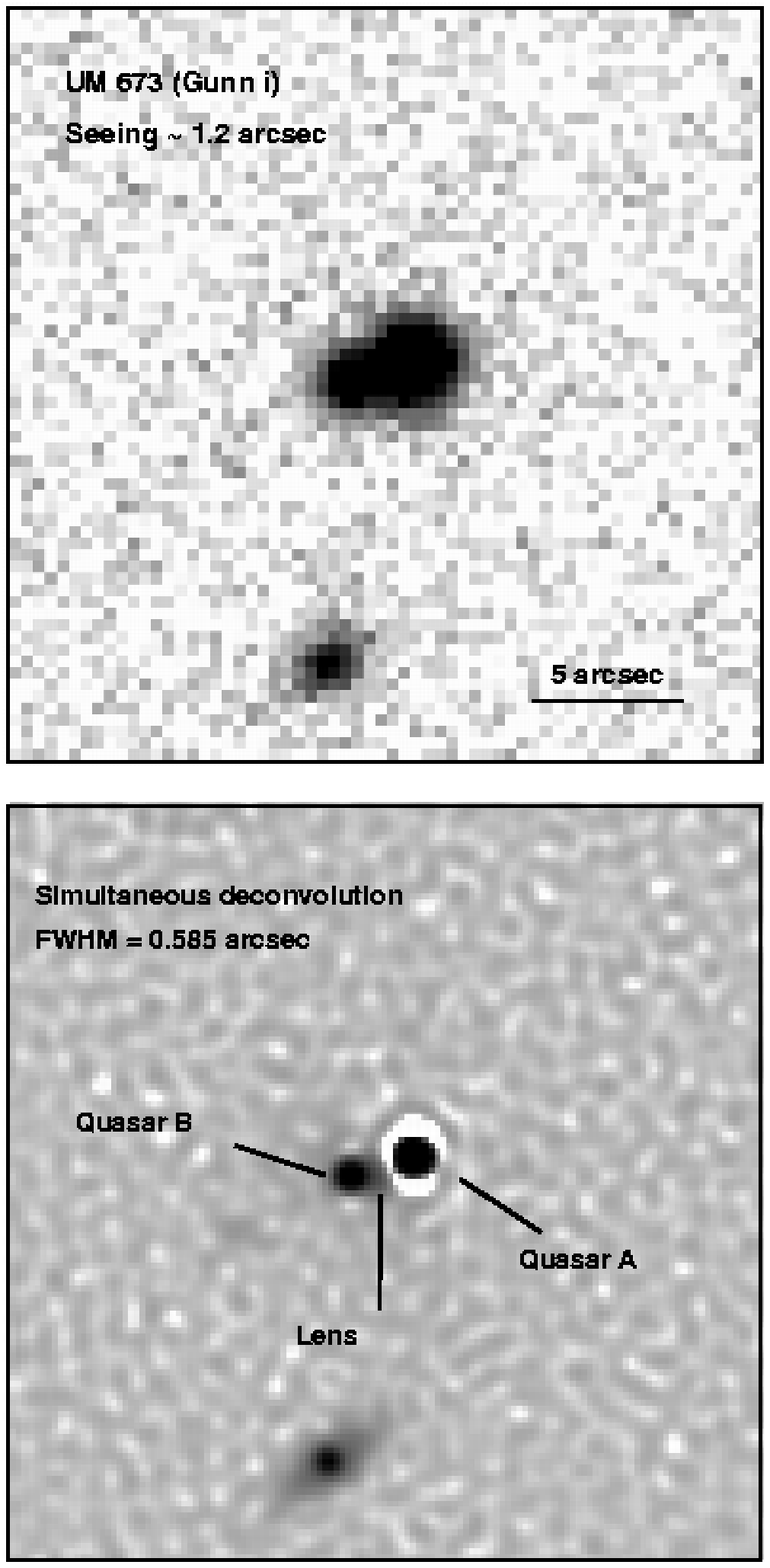}
\caption{
  {\it  Upper panel:} a $25\arcsec  \times 25\arcsec$ region extracted
  from one   of the 63   $Gunn\,i$ CCD frames  of  \obj.   The  seeing  is
  1.2$\arcsec$ and the   pixel scale is  0.39\arcsec. {\it  Lower panel:}
  simultaneous deconvolution of the 63 frames, with a final resolution
  of 0.585$\arcsec$  and  a  pixel  size  of 0.195\arcsec.  Due  to  the PSF
  variations in the DFOSC field,  minor artifacts were produced in the
  vicinity of the brighter quasar image.  They represent only 0.5\% of
  its peak intensity. The  lensing galaxy is marginally  seen between
  the two QSO components, close to component  B.  North is up, East is
  to the left.  A logarithmic intensity scale has been used to display
  the images.}
\label{in-dec}
\end{figure}

%----------------------------------------------------
\section{Observations}

The observations of \obj\ were carried out on a weekly basis using
the Danish 1.54\,m telescope at La Silla (Chile); at least three CCD frames
were taken each night for each  filter.  In total,  63
CCD frames were obtained    in the $Gunn\,i$  filter,  corresponding   to
18 observing nights, from October 3,  1998 to September  29, 1999. In
the Johnson $V$-band, 91 CCD frames were obtained, corresponding  to 23
nights.  18 out of these 23 $V$-band  epochs were taken during the 
same nights as the 18 $Gunn\,i$  CCD frames.  The remaining 5 
$V$-band   epochs were obtained between August 29 to December 15,
2001.

The CCD mounted on   DFOSC (Danish Faint Object  Spectrograph  Camera)
until  September 2000  was  a backside  illuminated LORAL/LESSER chip,
with a pixel  scale of 0\farcs39 pixel$^{-1}$  and a field of  view of
$13.7\times 13.7$  arcmin$^2$.  This detector was  then replaced by an
EEV/MAT chip, with  the same pixel  scale and FOV.  The exposure  time
for  the individual  CCD frame ($>3$ per  epoch) in  $Gunn\,i$  was 250
seconds.   For the  $V$-band data, the  integration  time  was varying
between 150  and 600 seconds  depending on the  moon light and seeing,
with  50\% of the frames having  an exposure time  of 250 seconds.  No
data were  lost due to  bad weather conditions, thanks to the
flexible   scheduling    applied    during  this  observing   program. 
Fig.~\ref{findingchart}   shows a part  of a  typical $Gunn\,i$ CCD frame. 
\obj\ can  be identified  on  the north-east part of  the  field.  Two
neighboring stars, labeled S1 and  S2, extensively used during the
data reduction, are also indicated.

\section{Data reduction techniques}

We   performed the photometric     reduction and analysis  using three
different  numerical methods to treat with photometry of blended
objects.   All three methods    were  developed at the  Institute   of
Astrophysics and Geophysics of  the University of Li\`ege.   The first
one, the  so-called  \emph{MCS},  is based on   deconvolution (Magain,
Courbin \& Sohy~\cite{magain98}).   It  has been extensively  used  to
obtain  the light-curves of many   lensed quasars (e.g.  Burud et  al. 
2002a,b).  The second method, \emph{Differential Imaging}, is based on
a matching  kernel technique similar  to that developed by Phillips \&
Davis~(\cite{phillips95}). The last  method, \emph{General}, is a
profile   fitting technique that  has  been  used to  analyse data of
lensed quasars,    e.g.,   by Remy~(\cite{remy96})  and   Courbin   et
al.~(\cite{courbin95}).

%----------------------------------------------------

\begin{figure*}[p]                                          
\centering
\includegraphics[width=11cm]{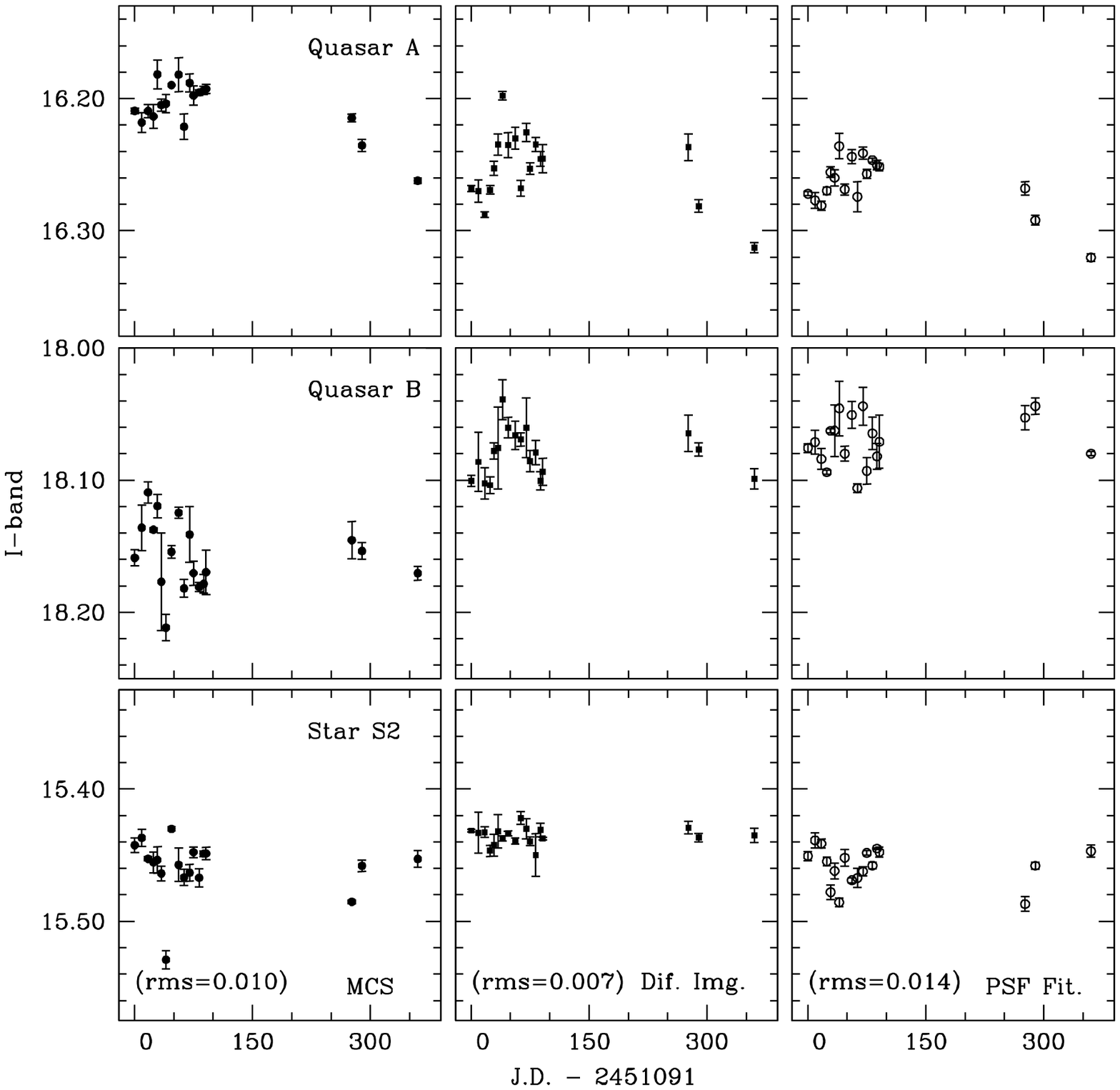}    
\caption{Cousins  $I$-band  light-curves,  for  component  A  (top),
  component  B (middle),  and for  the  reference star S2 (bottom), as
  obtained  from left to  right with 1- the  MCS deconvolution, 2- the
  Differential Imaging  method,  and 3-  the  PSF fitting method.  The
  dispersion between the photometric points of star S2 is indicated in
  each panel.   The  Julian  Date   2\,451\,091 corresponds to   1998,
  October 3.}
\label{I-lc} 

\centering
\includegraphics[width=11cm]{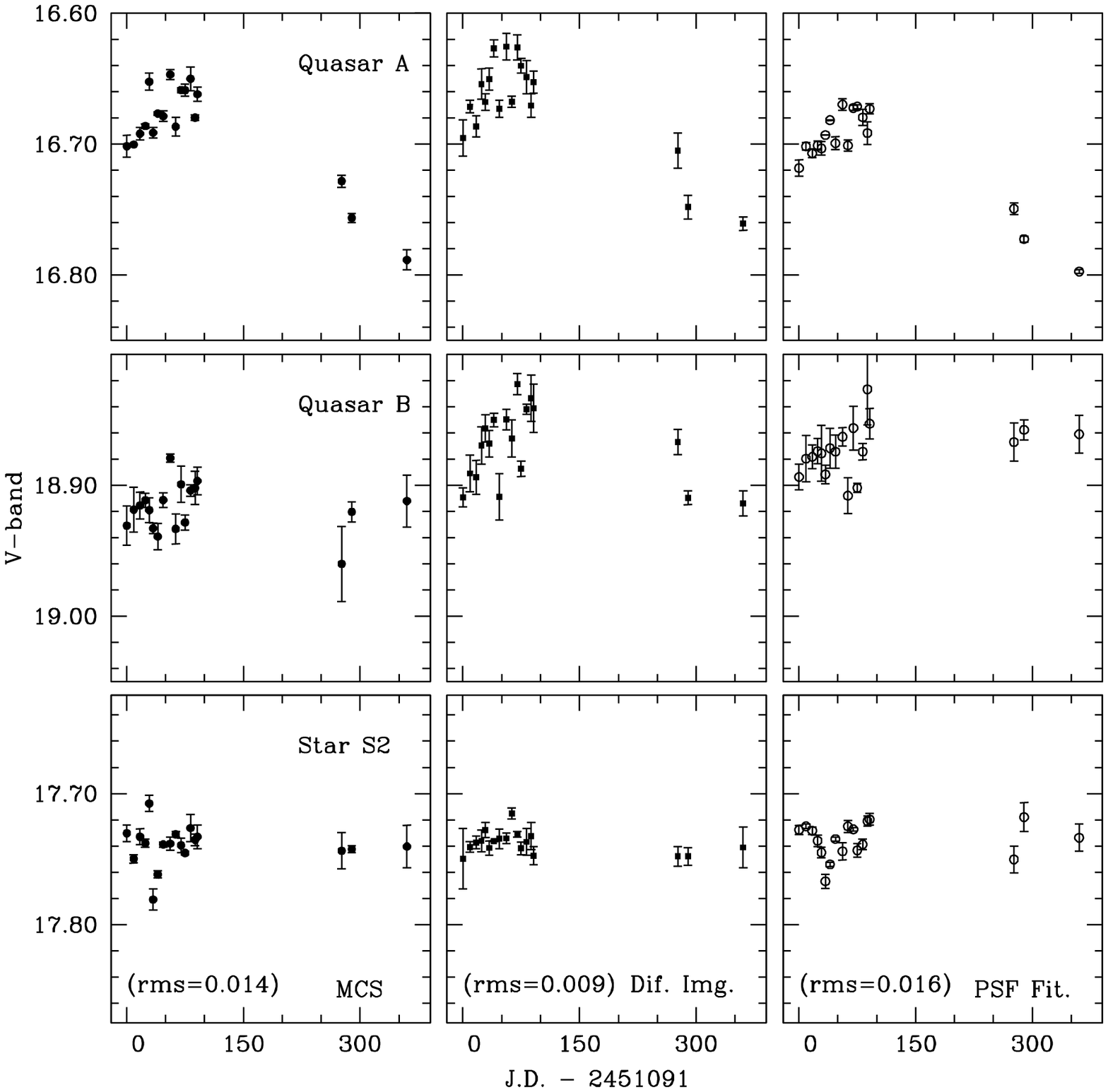}    
\caption{
  Cousins  $V$-band light-curves, spanning a  period of 400 days (same
  symbols and organization as in Fig.~\ref{I-lc}). Each point 
  corresponds to observations taken a few minutes before or after 
  the $I$-band data shown in Fig.~\ref{I-lc}.}
\label{V-lc-400d} 
\end{figure*}

%----------------------------------------------------

Prior    to  the  photometric   analysis    itself,  all frames   were
pre-processed, i.e. bias-subtracted, flat-fielded and rescaled to the
same    pixel-coordinate  reference     system,  using standard   
IRAF\footnote{Image Reduction and  Analysis  Facility --- is  written  
and supported  at the  National Optical Astronomy  Observatories (NOAO)}
procedures. Sky subtraction was performed using SExtractor (Bertin
~\cite{bertin96}), which can treat complicated sky structures, by
removing the   objects  from  the  data and   by  estimating the   sky
background through a  grid of cells adapted to  the object density  of
the field.  In the  case of  UM~673, the  process  resulted in a  very
accurate sky subtraction, at the percent level, due to the low stellar
density and  to the small  spatial variations of the  sky value in the
original data.

The PSF varies  significantly across the  DFOSC field of  view and the
observed variation changes from one  frame to another. Because of  the
very  low number of stars  suitable for the  PSF determination, it was
not possible  to properly  take  this  effect  into account.  This  is
probably the major cause of photometric uncertainties in our results.

\subsection{Image deconvolution}

The  main  difference     between   the   MCS  (Magain,   Courbin  
\& Sohy~\cite{magain98})   method and most  others   is the 
simultaneous deconvolution capability of   the algorithm: the
deconvolved  image is computed by minimizing the $\chi^2$ between 
all the individual frames and  a unique model image.    The
individual $\chi^2$ corresponding to each  image are  minimized at
the  same  time, hence the name  of \emph{simultaneous
deconvolution}.  All the available data are involved in this
simultaneous fit, including those  taken under mediocre observing
conditions.

The  deconvolved  image  is   decomposed  into  a  sum   of
analytical (Gaussian) point sources and a  numerical deconvolved
component  which represents extended objects, such as the lensing
galaxy or even the quasar  host   galaxy, if  detected   at  all.  
In  addition  to  the deconvolved image, the procedure returns the
peak intensity and  astrometric center of all point sources.   The
intensities are allowed  to vary  from one image to the next, hence
leading to  light-curves. The position of the point sources is
forced  to be the same  in all images. The astrometry of the images 
with poor seeing  is therefore constrained  by the ones with better
seeing.

The signal-to-noise of the final image is  thus constrained by the 
whole dataset. It also  has  an improved spatial  resolution, sampling
and depth.  Fig.  \ref{in-dec}  compares a simultaneously  deconvolved
image with  one of the original  data, in the  $Gunn\,i$  filter.  The
chosen   resolution is 0.585$\arcsec$ and  the   improved pixel size is
0.195\arcsec.  The ``ring'' at the wings of  the brighter component is
due   to   strong PSF variations    in  the   DFOSC  field  (see  also
section~\ref{DifIm}). Its flux  corresponds to $\approx$ 0.5\% of
component's A peak flux, for the $Gunn\,i$-band. For the $V$-band
this effect is weaker, of the order of 0.1\%.

Four  stars in the  vicinity of  \obj\ were  used to construct the
PSF. A $64\times64$ pixel sub-frame, centered on UM\,673, is extracted
from each DFOSC   frame.  Prior to  deconvolution, this  sub-frame is
flux-normalized,  by  division  of  the  integrated   flux of    the
non-variable star S1  (see Fig.~\ref{findingchart}).  This cancels the
effect of variable airmass and sky transparency.  Star S2 is used as
a comparison, in order to check the relative photometry, to control the
errors and  to quantify the  effect of the  PSF variations across the
field.   Both stars  have been  found   to be photometrically   stable
(Sinachopoulos et  al.~\cite{sina01};    Nakos et al.~\cite{nakos03}). 
Unfortunately, no  other  bright, isolated  star is  present in  the
field of view.

The  $Gunn\,i$ and   $V$  fluxes of   the  two  quasar  components are
converted to  standard Johnson-Cousins magnitudes using the literature
values of the reference   star S2 (Nakos  et  al.~\cite{nakos03}).  We
notice  a slight shift between the   $V$ magnitude of  S2  and the one
published  in the literature,   probably  due  to differences  in  the
apertures used to carry out the flux normalization relative to S1. The
corrected value of the  Cousins magnitude of S2  is 0.04 fainter than
in Nakos  et al.~(\cite{nakos03}), i.e.   $V =17.74 \pm 0.02$  and
$(V-I) =2.28$.

%----------------------------------------------------

\begin{figure*}[t!]                                           
\centering \includegraphics[width=13cm]{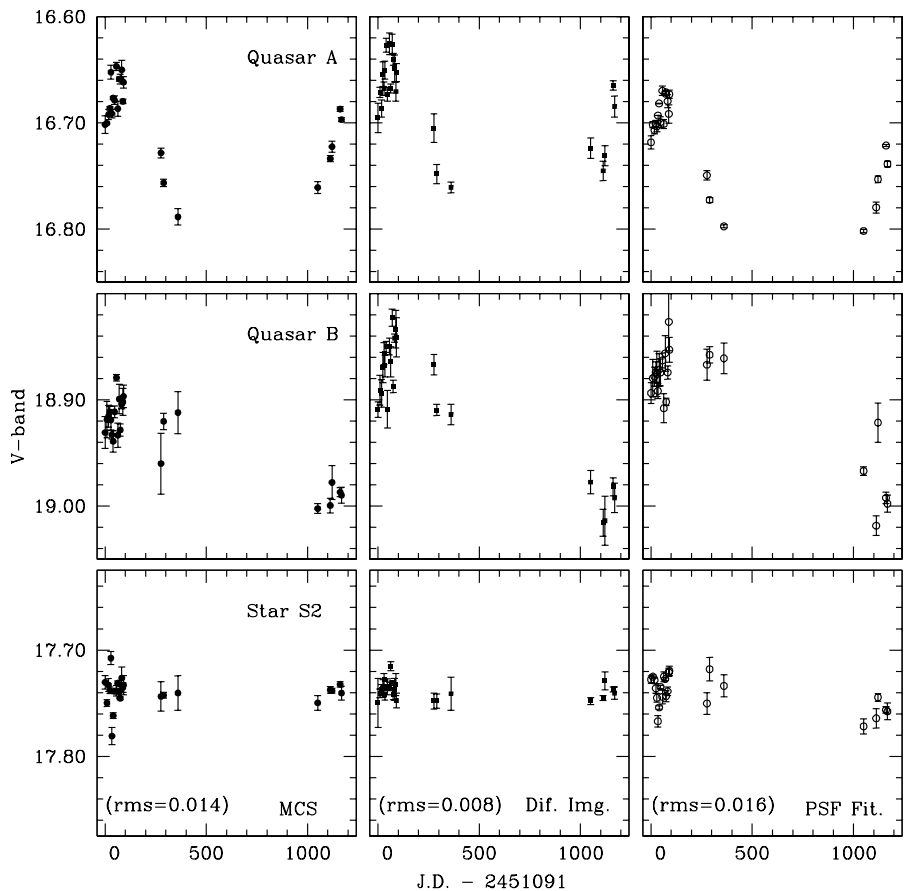}    
\caption{
Observations in  the $V$-band were carried out  for a longer time
than  the $Gunn\,i$ ones,  extending the light-curve up  to 1200 days
(same symbols and organization as in Fig.~\ref{I-lc}).  }
\label{V-lc-1200d}  \end{figure*}

%----------------------------------------------------

\subsection{\label{DifIm} Differential imaging}

The  basic   concept  of  this  method,  implemented  in    a new code
(Python/IRAF/C++),   contains   elements     taken  from    the   IRAF
task {\it  psfmatch}  (Phillips  \&  Davis~\cite{phillips95}).  It  is
based on the comparison of a set of images, which have been registered
and transformed so that their PSF and photometry perfectly match.

First, the best seeing and S/N image is identified  and labeled as the
reference image, all the other frames  being called {\it input frames}
in the following.  All the frames are rescaled so that star S1  has a  
constant flux  in  all images. 

The reference frame is subsequently  degraded to match the resolution
of the  other frames.  This is  done by convolving the reference frame
with a transformation  kernel  computed in  the Fourier   space.  Each
frame is then subtracted from  the degraded reference, hence providing
a residual  frame showing only   relative flux variations.   Following
this procedure, non-variable objects are removed, i.e. in the present
case the contamination by the lensing galaxy is canceled.

Importantly, while  this  method   provides  accurate relative
photometry,  it   does not give   any   information on the  absolute
photometric zero-point.  This information  is obtained by using the 
photometry from the PSF fitting method (see  section~\ref{psffit}) as
a  cross calibration. Since the PSF fitting does not take the
lensing   galaxy into  account, slight shifts  are possible   on the
absolute  scales   adopted for   the light-curves produced with  MCS
and with the differential  imaging or PSF fitting method.

It is possible in the implementation of the code to use a convolution
kernel that  changes across the  field of view.   However, due to the
very low  number of useful PSF  stars in the  field of  \obj, a fixed
kernel was  used in the  present application.  Trial and error showed
that the best kernel  was obtained when star S1 alone  was used for the
computation.  It is indeed the bright star closest to \obj, for which the
PSF variations are expected to be the smallest.

%----------------------------------------------------

\begin{figure*}[t!]                                           
\centering
\includegraphics[width=8.75cm]{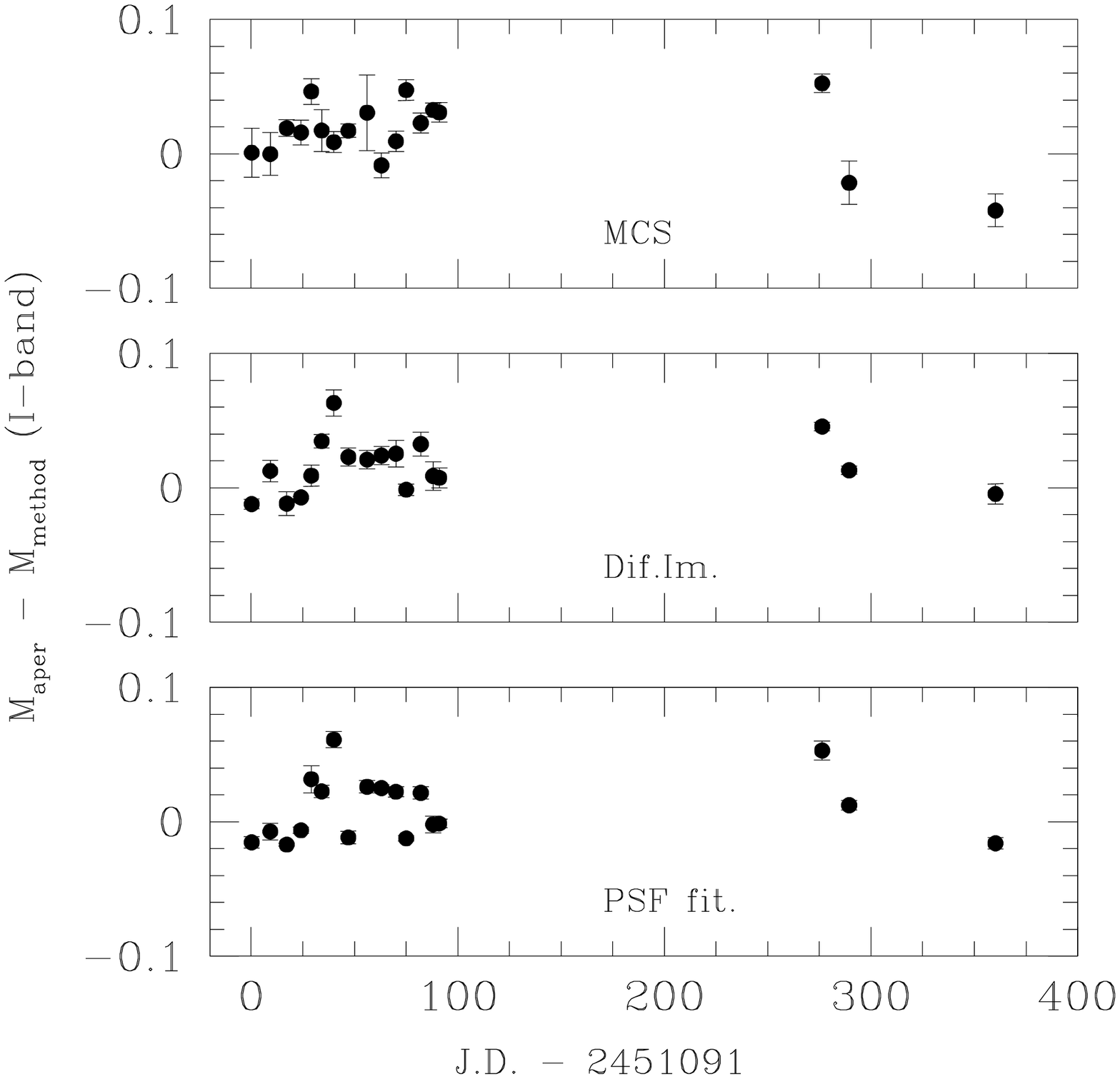}   
\includegraphics[width=8.75cm]{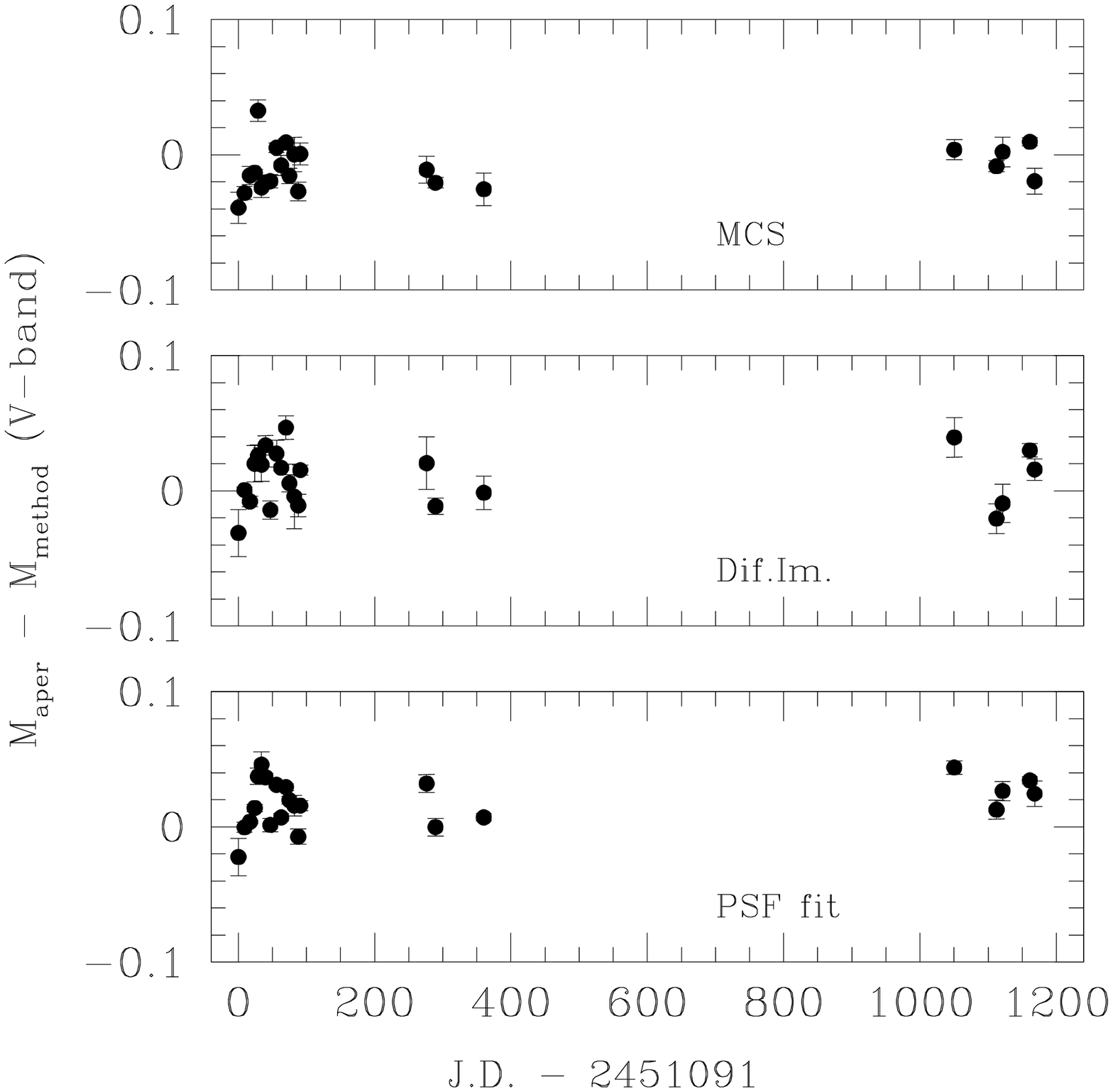}  
\caption{
  {\it   Left Panel:} difference    $I$-band light-curve  between  the
  aperture  photometry and the  magnitude corresponding to  the sum of
  the  fluxes  of  the two  QSO   components (A+B).  The  upper  panel
  corresponds  to the MCS   algorithm,  the middle   panel is for  the
  Differential  Imaging, and  the lower  panel is for  the PSF Fitting
  method.   {\it Right Panel:}  idem for  the  $V$-band.}
\label{Dmag-I} 
\end{figure*}

%----------------------------------------------------

\subsection{PSF fitting\label{psffit}}

``General''   is a profile fitting  technique  that has already been
applied to  several   gravitational  lens  systems  (e.g.,  Burud   et
al.~\cite{burud98};  {\O}stensen et al.~\cite{ostensen97}; Courbin
et     al.~\cite{courbin95}).  A set of field     stars  is used to
construct   a {\it numerical}  profile to  be fitted  to the quasar
components, together  with     a set of    analytical profiles  that
represent extended objects.   The advantage of this  method is that it
has few free parameters: the relative  position and intensities of the
quasar components, and the shape  parameters of the lens (ellipticity,
position  angle).  All parameters   of the point  sources and  lensing
galaxy are fitted simultaneously to the data.

The stars used to  build the PSF for  each frame are the  ones
labeled S1-S4  in    Sinachopoulos    et al.~(\cite{sina01})  and
Nakos  et al.~(\cite{nakos03}). As for the other two methods, the 
flux of star S1 is taken as a reference to normalize all epochs.

The  application of  the  method to   the  present data   set  was not
straightforward, due  to  the large pixel size   of the detector that 
hampers the accurate determination  of the lens position. We therefore
model \obj\ as two  point sources only.  The  consequence is that  the
flux we measure for component B is contaminated by the lensing galaxy,
even though the  residuals after the simultaneous  fit of the two PSFs
to   the quasar images are  good.  However, even  with this contamination, the
total summed fluxes  of   the two quasar images remain compatible   
with the flux measured  through a large aperture  applied  directly to 
the data (see Fig.~\ref{Dmag-I}).

%----------------------------------------------------

\begin{figure*}[t!]                                            
\centering
\includegraphics[width=13.0cm]{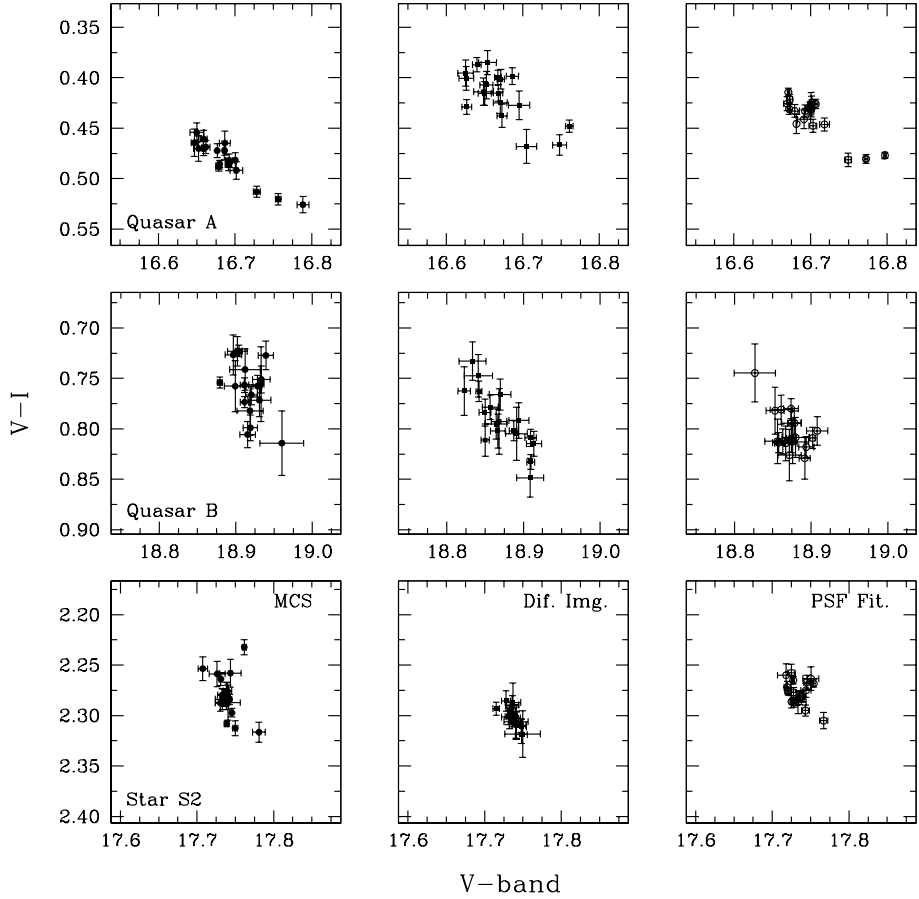}          
\caption{Color (V-I)  dependence  of   the   photometric  variations,   for
  components A, B and for the reference star  S2, as a function of the
  Johnson $V$  magnitudes. A clear  trend  is visible for  the  quasar
  image A, becoming bluer when it is getting brighter.}
\label{VmI_versus_V} 
\end{figure*}

%----------------------------------------------------

\section{Results}

The  photometry of UM~673A  and B,  and of  the reference  star S2, is
displayed   in  Figs.~\ref{I-lc}-\ref{V-lc-1200d}.   Between 3  and  6
frames were obtained  for  each observing  epoch and each  filter. The
values plotted in the figures are the mean values for each night.  The
1$\sigma$ error bar for each epoch is taken  as the error on this mean
value, i.e. the standard deviation between the magnitudes measured in the
individual images at a given epoch, divided by $\sqrt N$, where $N$ is
the  number of frames.   The systematic errors  due  to the zero-point
corrections,  which are   of the  order  of  0.02  magnitude, are  not
incorporated in the error bars.

On short time-scales, where both $V$ and $I$-band data are available
(see Figs.~\ref{I-lc}  \&   \ref{V-lc-400d}), all  methods  agree that
component A shows  a peak-to-peak variability  of about 0.08 magnitude
in  the $I$-band,  and about 0.14  magnitude  in the Johnson $V$-band. 
The  time scale of the variability  is of the  order  of 350 days, for
both filters.   For  component   B, where   the uncertainty    in  the
photometry is larger, the short time-scale variability is at the limit
of  detection, for both  filters.    However, the  five points  around
JD$\sim$1100, placed more than $3\sigma$  away from the points between
JD$\sim$0 and JD$\sim$350, clearly show  that, on larger  time-scales,
component   B  has   also    undergone  photometric variations    (see
Fig.~\ref{V-lc-1200d}).

The  comparison star S2  is   used as an   independent  check of  each
photometric  method.  It is  found  that the  dispersions between  the
photometric points (i.e.  between all the epochs,  as shown in  Figs. 
3-5) are compatible with  the error   bars  estimated for  each
photometric method.

\section{Discussion}

\subsection{Comparison of the methods}

We have analyzed a two-band  photometric data set of the gravitational
lens system  \obj\ (18 epochs in $Gunn\,i$,  23 epochs for the Johnson
V-band),  using  three different  photometric  methods.   All  methods
clearly  reveal   photometric  variations,  both    over short   (i.e. 
$\approx$ 350 days) and long (i.e.  $\approx$ 1200 days) time scales.
The variability of the  brighter component A  is well measured  by all
three methods, but is not as clear for the fainter image B.

The three photometric methods are applied to the data independently
of any prior knowledge on the geometry of \obj, e.g., as can be done
from HST images (Leh\'ar et al.~\cite{lehar00}).
This	choice allows us   to estimate   the 
relative  merits and  drawbacks   of  the methods,  especially  with
respect to the  systematic errors, due  to the PSF variations or  to
the design of the methods themselves.  The only internal calibration
applied  is that  the absolute  flux ratio of  the quasar images, as
derived from the differential imaging, is determined by applying the
absolute flux ratio from the PSF fitting method.  The lensing galaxy
is not taken into account in  the PSF fitting. Our cross-calibration
of one method using  the other  has  no consequences on  the $V$-band
light-curves where  the lens is faint. However, it has an effect on
the $I$-band light-curves: Fig.~\ref{I-lc} shows a systematic offset
between the   QSO light-curves obtained with  MCS  and the other two
methods.   The shift is in the  opposite direction  between images A
and B.  This effect  is not present in  the curves of  the reference
star  S2 and is probably due to the contamination  of the lens affecting the
different methods in different ways.  A zero-point miscalculation is
excluded, since this would have resulted in the two curves moving in
the same direction.

PSF variation or distortion does not have the  same effect on the
different methods.   An outlier is seen, for example, in the $I$-band
light-curve of star S2 when using the MCS method ($I = 15.53$), and
corresponds to a set of images with bad focusing.  This point  is also an
outlier   in the  PSF	fitting light-curve   and reveals   a higher
sensitivity of these methods to imperfect PSF determination.

The systematic errors are consistent among the methods,  but show a
slight filter-dependence. In Fig.~\ref{Dmag-I}  we  present the
difference curve between the total light,  as measured by  direct
aperture photometry of \obj, and the sum of the fluxes of the two 
QSO components, as measured by the three methods.  For the $I$-band
(left panel),  the points lie systematically above the mean value
$<\Delta mag>\, = \,0,$ indicating that all methods overestimate
the sum of the fluxes (A+B)  at the level of 2\%. For the $V$-band
however (right panel),  the spread of the points around the mean
value $<\Delta mag>\, = \,0$ is more homogeneous, leaving a hint
that the presence of the galaxy in the $I$-band has an effect,
although minor, in the  photometric results obtained by the three
methods.

\subsection{The time delay}

The parts of the light-curves that are  the best sampled correspond to
the  15 first measurements, spanning  120  days. Considering only this
part of the curves and   neglecting possible microlensing events,  we
have attempted to  estimate the time delay  between the quasar images. 
Unfortunately the peak-to-peak amplitude  of the photometric variation
is only 0.08  magnitude  in the $V$-band  and  0.05 magnitude in  $I$,
making it impossible to estimate the time delay. This is also the reason
why  it was decided to  stop the observations  of  \obj\ after 3
months of   monitoring, giving priority  to   other, more variable
objects. 

\obj\ shows larger photometric variations over longer time-scales, but
these  variations do not   match at all  after  shifting the  curve of
component B  relative to that  of component  A.  These  long  term
variations are sampled, in the $V$-band  light-curves, with only three
points at JD$\approx$350 days  and with five points at JD$\approx$1100
days. The two groups of points span 150 days  each.  The fact that the
light-curves do not match  on long time  scales has two possible
interpretations: {\bf 1-} the time delay is longer  than the length of
the groups of points at  JD$\approx$350 days and JD$\approx$1100 days,
i.e. $\Delta t >$  150 days or, {\bf  2-} the variations are  largely
dominated by microlensing.  We will see in the following that there is
possible evidence supporting the latter interpretation rather than the
former one, also given that predicted time delays for \obj\ are of the
order of a few months (Surdej et al.~\cite{surdej88}).

\subsection{Evidence for microlensing}

Although the large gaps in the light-curves of \obj\  do not allow the
measurement of  the time delay,   the photometric variations  observed
over long time scales are indicative of chromatic microlensing.

  We have plotted in  Fig.~\ref{VmI_versus_V}  the $V-I$  color index as a
  function of the $V$  magnitude, for the two   quasar images and  for
  star S2, as derived  with the three  photometric methods.  Since the
  error bars on the $V$-band photometry  are comparable with the error
  bars on   the   $I$-band points,  the  distributions  of  points  in
  Fig.~\ref{VmI_versus_V} should   be  elliptical, with   a  long-axis
  parallel to    the y-axis and    $\sqrt 2$  times  larger   than the
  (horizontal) short axis, in the case of no correlation between color
  and magnitude.   While this is  the observed situation for  star S2,
  the   plot for component   A of the quasar   displays a clear linear
  trend, indicating that it is becoming  bluer as it gets brighter, in
  agreement  with   microlensing   scenarios   where  a compact  quasar
  accretion   disk   is (micro)lensed   by    a  network of  caustics. 
  Microlensing  preferentially amplifies the  most central and  bluer
  parts of the source with respect to  more external and redder parts;
  an effect that can  be seen in the  continuum of a quasar spectrum
  (e.g.       Wisotzki      et  al.~\cite{wiso93};      Courbin     et
  al.~\cite{courbin00}).

This result is in apparent contradiction to the results of Wisotzki et al. 
~(\cite{wiso04}), who exclude the microlensing scenario from integral
field spectroscopy,  but their  observations  were performed  in 2002. 
This is three years after the latest of the epochs of our light-curves
where both $V$ and $I$-band observations are available.

The reliability of the trend we observe in Fig.~\ref{VmI_versus_V} can
be tested by comparing the V-I index obtained using one method, to the
index measured  by using the  other  two methods.   If all methods are
sensitive in the same way to a real  $V-I$ trend as  a function of the
$V$ magnitude, then the $V-I$ colors measured by two different methods
should  correlate  well.   If  there  is no   real  trend,  the  $V-I$
measurements are dominated by photon noise,  and no correlation should
be found.   Doing this test reveals  a  strong correlation between the
results in the three methods for quasar A, but not for quasar B or for
the reference star  S2.   The color-dependent variations seen   in the
quasar image A can therefore be  safely considered as real.

Although  the case of microlensing being   present for both components
would  be rather unusual,  the    color  dependence on the    $V$-band
magnitude    for component   B,     shown  by  differential    imaging
(Fig.~\ref{VmI_versus_V}), is    quite       puzzling   and   requires
confirmation.    Given the fact that  none  of the   two other methods
confirms the above   trend, either higher   signal-to-noise multicolor
images are   required,  or an  extensive   high SNR spectrophotometric
monitoring of UM~673.

If the variation we have observed is not due to microlensing,
the peak-to-peak  variation detected  over  three years  is only  0.14
magnitude in the  $V$-band.  The photometric  error on the individual
points is about 0.01  magnitude  for all  three  methods used in  this
paper and the visibility of UM~673 is about  6 months across the year. 
According to the simulations by Eigenbrod et al.~(\cite{eigen05}), the
minimum sampling to  adopt in order  to derive the time-delay  with an
accuracy better than 2\%  in less than 2  years is of one point every
three days,  making  \obj\ a very  ``expensive''  object  in terms  of
telescope time,  if the goal  is to use  it to determine H$_0$.  Given
the  chromatic   variations   detected in  the   present   photometric
monitoring, \obj\ may however turn out to be a very interesting object
for microlensing studies.

\begin{acknowledgements} 
  
  We   thank IJAF and   ESO for granting  us  observing  time for this
  project  on a flexible  basis.  Th.N.  acknowledges support from the
  project  ``Chercheurs  Suppl\'ementaires    aux     \'Etablissements
  Scientifiques F\'ed\'eraux". Part of the research was also performed
  in   the  framework of  the  IUAP P5/36  project,   supported by the
  DWTC/SSTC  Belgian  Federal  services.   Th.N. also  acknowledges C. 
  Abajas  Bustillo  for the  interesting discussions.  FC acknowledges
  financial support    P\^ole  d'Attraction  Interuniversitaire, P4/05
  (OSTC, Belgium). J.S. wishes to acknowledge support from the Belgian
  OSTC PRODEX program ``Gravitational Lensing".
  
\end{acknowledgements}

%--------------------------------------------
%--------------------------------------------

\end{document}